\documentstyle[aps,preprint,tighten]{revtex}

\newcommand{\be}{\begin{equation}}
\newcommand{\ee}{\end{equation}}
\newcommand{\beq}{\begin{eqnarray}}
\newcommand{\eeq}{\end{eqnarray}}

\newcommand{\eps}{\varepsilon}

\newcommand{\tE}{\lefteqn{\smash{\mathop{\vphantom{<}}\limits^{\;\sim}}}E}
\newcommand{\tP}{\lefteqn{\smash{\mathop{\vphantom{<}}\limits^{\;\sim}}}P}
\newcommand{\tQ}{\lefteqn{\smash{\mathop{\vphantom{<}}\limits^{\;\sim}}}Q}
\newcommand{\Et}{\lefteqn{\smash{\mathop{\vphantom{\Bigl(}}\limits_{\sim}
\atop \ }}E}
\newcommand{\Pt}{\lefteqn{\smash{\mathop{\vphantom{\Bigl(}}\limits_{\sim}
\atop \ }}P}
\newcommand{\Qt}{\lefteqn{\smash{\mathop{\vphantom{\Bigl(}}\limits_{\sim}
\atop \ }}Q}

\newcommand{\tN}{\lefteqn{\mathop{\vphantom{\Bigl(}}\limits_{\sim}
\atop \ }N}
\newcommand{\tM}{\lefteqn{\mathop{\vphantom{\Bigl(}}\limits_{\,\sim}
\atop \ }M}
\newcommand{\tNn}{\lefteqn{\mathop{\vphantom{\Bigl(}}\limits_{\sim}
\atop \ }{\cal N}}

\newcommand{\R}{R}
\newcommand{\SA}{{\cal A}}
\newcommand{\tPb}{{\tP_{\smash{(\beta)}}}}

\newcommand{\nd}{{\cal N}_D}

\newcommand{\G}{{\cal G}}

\begin{document}

\title{ \Large \bf 
Area spectrum in Lorentz covariant loop gravity}
\author{\normalsize S.~Alexandrov$^{1,2,4,}$\thanks{e.mail: 
alexand@spht.saclay.cea.fr} and
D.~Vassilevich$^{3,4,}$\thanks{e.mail: 
vassil@itp.uni-leipzig.de}}

\address{{$^1$
Service de Physique Th\'eorique, C.E.A. - Saclay, 91191 Gif-sur-Yvette
CEDEX, France}\\
$^2$Laboratoire de Physique Th\'eorique de l'\'Ecole Normale
Sup\'erieure, 24 rue Lhomond, 75231 Paris Cedex 05, France\\
{$^3$Institute for Theoretical Physics, Leipzig
University, Augustusplatz 10/11, 04109 Leipzig, Germany}\\
{$^4$ V.A.~Fock Department of 
Theoretical Physics, St.~Petersburg
University, Russia}}

\maketitle

\begin{abstract}
We use the manifestly Lorentz covariant canonical formalism to
evaluate eigenvalues of the area operator acting on Wilson lines.
To this end we modify the standard definition of the loop states
to make it applicable to the present case of non-commutative
connections. The area operator is diagonalized by using the
usual shift ambiguity in definition of the connection. The eigenvalues
are then expressed through quadratic Casimir operators. No dependence
on the Immirzi parameter appears. 
\end{abstract}

\section{Introduction}

Quantization of gravity is an extremely hard and interesting problem 
which remains unsolved so far. During last years a number of 
approaches have achieved a definite progress in treating various 
aspects of quantum gravity. The most elaborated and popular line 
of research is string theory which includes perturbative gravity in
its spectrum and unifies it with other interactions. 
An alternative (or, perhaps, complementary) approach is the loop quantum
gravity \cite{loop} (for review, see \cite{Rov-dif}). 
This program relies on the Dirac canonical quantization. 
It is explicitly nonperturbative and 
background independent so realizing the basic principles of general
relativity. During the previous decade this approach has got rigorous 
mathematical foundations \cite{cyl} and has led to interesting qualitative
predictions about quantum spacetime. 
    
These predictions originate from remarkable results
obtained in the framework of loop quantum gravity, which are calculations
of the volume and area spectra \cite{area}-\cite{volume}. 
It appeared, however,
that the area spectrum depends on the so called Immirzi parameter \cite{imir}.
It parametrises a canonical transformation \cite{barb}
which introduces a new connection field. 
The reason for 
this dependence is that this transformation cannot be
realized unitarily in the Hilbert space of quantum theory \cite{Rov-Tim}.
In the language of quantum field theory this means presence
of a quantum anomaly. There exist two different types of the quantum
anomalies. The first type of the anomalies appear when a symmetry
of the classical action cannot be preserved by quantization
due to divergences or other quantum effects. Chiral and conformal
anomalies belong to this type. Their presence indicates emergence
of a new physics. The most celebrated example is the chiral anomaly
in QCD which has been used for description of the low energy hadron
physics since late 60's. Rather naturally, it has been suggested 
\cite{Rov-Tim}
that the anomaly in the mentioned canonical transformation belongs
to this type and, consequently, the Immirzi parameter is a new
fundamental constant. 

One cannot however exclude the second possibility. An anomaly could
appear if a symmetry is involuntary broken by the choice of a particular
quantization scheme. If this is the case, the remedy can be in
applying another quantization scheme which explicitly preserves
as much important symmetries as possible. This is the rout we take
in the present paper by applying the manifestly
Lorentz covariant quantization of \cite{SA} to calculation of the
area spectrum.

There are already some evidences that the Immirzi parameter
dependence may disappear in a more symmetric quantization scheme.
In the paper \cite{SA} the path integral quantization scheme
of \cite{AV} has been extended to arbitrary values of the Immirzi
parameter. It has been demonstrated, that the Immirzi parameter
dependence does not appear in the path integral. We should stress,
that in principle the path integral formalism is capable to see
non-perturbative effects (as e.g. the virtual black hole formation
\cite{KuVa}). Another important result was obtained recently
by Samuel \cite{Sam} who demonstrated that the Barbero connection
is not a Lorentz connection.

Recently, the importance for the theory to be Lorentz-covariant has
been also recognized in spin foam models \cite{sfLor} which represent
the modern development of loop quantum gravity \cite{sf}. 
However, the Lorentz covariance has been
introduced there without any reference to the canonical quantization.  
It is an important task to develop a Lorentz covariant
formulation ``from the first principles''. 

In this paper we apply the Lorentz covariant canonical quantization
developed in \cite{SA} to loop quantum gravity. 
We re-derive the spectrum of the area operator in the new framework.
To this end we construct the Wilson line operator with true
Lorentz connection. Since the Dirac brackets of the connections
are non-zero, there is not connection representation. However,
by choosing an appropriate vacuum state we are able to construct
the quantum states corresponding to the Wilson lines which behave
in a very similar way to the ordinary loop states. However, the area
operator is not necessarily diagonal on these states. To diagonalize
this operator we use the usual ambiguity in the connection:
any connection can be shifted by a vector and will still remain
a proper connection. It appears, that the shift 
is uniquely defined by the requirements that it vanishes on the
constraint surface and
that the area operator is diagonal
on the Wilson line states. This new connection obeys a remarkably
simple bracket algebra. Eigenvalues of the area operator are then
calculated. They {\it do not} depend on the Immirzi parameter.

The paper is organized as follows. In the next section we summarize
the covariant canonical formulation of \cite{SA}. In sec. III we
discuss the choice of the connection variables to be used in the
Wilson line states. Area spectrum is calculated in sec. IV. 
Section V is devoted to discussion of the results, problems
and future perspectives. Appendices are intended to list various
definitions and useful properties.
 
We use the following notations for indices.
The indices
$i,j,\dots$ from the middle of the alphabet label the space coordinates.
The latin indices $a,b,\dots$ from the beginning of the alphabet
are the $so(3)$ indices, whereas
the capital letters $X,Y,\dots$ from the end of the alphabet are
the $so(3,1)$ indices.

\section{$SO(3,1)$-covariant canonical formulation}

In this section we review the covariant formalism developed in \cite{SA}.
It is a canonical formulation of general relativity based on 
the generalized Hilbert--Palatini action suggested by Holst \cite{Holst}
\be
S_{(\beta)}=\frac12 \int \eps_{\alpha\beta\gamma\delta}
e^\alpha \wedge e^\beta \wedge (\Omega^{\gamma\delta}+
\frac{1}{\beta}\star\Omega^{\gamma\delta}). \label{Sd1}
\ee
Here the star operator is defined as
$ \star \omega^{\alpha\beta}=
\frac 12 {\eps^{\alpha\beta}}_{\gamma\delta}
\omega^{\gamma\delta} $, and $\Omega^{\alpha\beta}$ is the curvature of
the spin-connection $\omega^{\alpha\beta}$.
A $3+1$ decomposition of the fields reads:
\beq
&e^0=Ndt+\chi_a E_i^a dx^i ,\quad e^a=E^a_idx^i+E^a_iN^idt,& \nonumber \\
&{\tE}^i_a =h^{1/2}E^i_a,  \quad
\tN=h^{-1/2} N, \quad \sqrt{h}=\det E^a_i,& \label{tetrad} \\
&N^i=\nd^i+\tE^i_a\chi^a\tNn, \quad
\tN=\tNn+\Et_i^a\chi_a\nd^i.& \nonumber
\eeq
Here $E^i_a$ is the inverse of $E_i^a$.
The field $\chi_a$ describes deviation
of the normal to the spacelike hypersurface $\{ t=0\}$ from
the time direction.

Let us introduce matrix fields carrying one Lorentz index
 \beq
 & A^{ X}=(\frac12 \omega^{0a},\frac12 {\eps^a}_{bc}\omega^{bc})
     &{\rm -\ connection\ multiplet}, \nonumber\\
&  \tP_X^{ i}=(\tE^i_a,{\eps_a}^{bc}\tE^i_b\chi_c)
     &{\rm -\ first\ triad\ multiplet}, \label{multHP}\\
&  \tQ_X^{ i}=(-{\eps_a}^{bc}\tE^i_b\chi_c,\tE^i_a)
     &{\rm -\ second\ triad\ multiplet}, \nonumber\\
&  \tPb_X^i=\tP_X^i-\frac{1}{\beta}\tQ_X^i
     &{\rm -\ canonical\ triad\ multiplet},  \nonumber
 \eeq
which form multiplets in the adjoint representation of so(3,1).
In Appendix A we present the relations between the triad
multiplets and introduce the numerical matrices  
$\Pi$ and $\R$ (\ref{p-q}), (\ref{pb-q})
appearing in the formulas below.  
In terms of these fields the decomposed action can be represented in the form:
\beq
S_{(\beta)} &=&\int dt\, d^3 x (\tPb^i_X\partial A^X_i
+{\cal N}_{\G}^X \G_X+\nd^i H_i+\tNn H),  \label{Sd3} \\
\G_X&=&\partial_i \tPb^i_X +f_{XY}^Z A^Y_i \tPb^i_Z,\nonumber \\
H_i&=&-\tPb^j_X F_{ij}^X , \nonumber \\
H&=&-\frac{1}{2\left(1+\frac{1}{\beta^2}\right)} 
\tPb^i_X \tPb^j_Y f^{XY}_Z R^Z_W F_{ij}^W, \nonumber \\
F^X_{ij}&=&\partial_i A_j^X-
\partial_j A_i^X+f_{YZ}^X A^Y_i A^Z_j, \nonumber
\eeq
where $f_{XY}^Z$ are $so(3,1)$ structure constants,
${\cal N}_{\G}^X=A_0^X$. The
$so(3,1)$ indices
are raised and lowered with the help of the Killing form 
\be
g_{XY}=\frac{1}{4} f_{XZ_1}^{Z_2}f_{YZ_2}^{Z_1},\quad
g^{XY}=(g^{-1})^{XY}, \quad
g_{XY} =\left(
\begin{array}{cc}
\delta_{ab}&0 \\ 0&-\delta_{ab}
\end{array}
\right).
\ee
The limit $\beta\to i$ gives Ashtekar gravity. Even though the
Hamiltonian constraint $H$ in (\ref{Sd3}) has apparently a pole
at $\beta =i$ one can demonstrate \cite{SA} that this limit
is non-singular.

The canonical variables of the model are $A_i^X$ and $\tPb^i_X$.
$G_X$, $H_i$ and $H$ are first class constraints
obeying the algebra presented in Appendix C. We call them
the Gauss law, diffeomorphism and Hamiltonian constraints respectively.
There are also two sets of the second class constraints.
They are represented by $3\times 3$ symmetric fields
\beq \phi^{ij}&=&\Pi^{XY}\tQ^i_X\tQ^j_Y=0, \label{phi} \\
\psi^{ij}&=&f^{XYZ}\tQ_X^{[l}\tQ_Y^{\{ j]}\partial_l \tQ_Z^{i\} }
-2(\tQ\tQ)^{\{ i[j\} }\tQ_Z^{l]}A_l^Z=0, \label{psi} \\
(\tQ\tQ)^{ij}&=&g^{XY}\tQ_X^i\tQ_Y^j.
\eeq
Symmetrization is taken with the weight $1/2$.
Antisymmetrization includes no weight.

The existence of the second class constraints gives rise to the Dirac 
bracket \cite{Dirac}
\be \{ K,L\}_D =\{ K,L\}-\{ K,\varphi_{r}\}(\Delta^{-1})^{rr'}
\{\varphi_{r'},L\}, \label{Dbr}  \ee
where $\varphi_r=(\phi^{ij},\psi^{ij})$.
The matrix of commutators of the second class constraints
$\Delta^{rr'}$ can be found in Appendix B.  
Both $\Delta$ and $\Delta^{-1}$ are triangular. Due to this when one of the 
functions in (\ref{Dbr}), $K$ or $L$
is a first class constraint, the Dirac bracket coincides
with the ordinary one (except for the case when $K=H$ and $L$
depends on the connection). In particular, this gives 
\beq
\{ \G_X,\G_Y\}_D&=&f_{XY}^Z \G_Z, \nonumber \\
\{ \G_X,A^Y_i\}_D&=&
\delta^Y_X\partial_i -f_{XZ}^Y A^Z_i,  \label{trans} \\
\{ \G_X,\tP_Y^i\}_D&=&f_{XY}^Z \tP_Z^i.  \nonumber
\eeq

Finally, the Dirac brackets of the canonical variables have the form:
\beq
\{ \tPb_X^i,\tPb_Y^j\}_D&=&0, \nonumber \\
\{ A^X_i,\tPb_Y^j\}_D&=&\delta_i^j\delta^X_Y-\frac12
\R^{XZ}\left(\tQ^j_Z\Qt_i^W+\delta^j_i I_{(Q)Z}^W
\right)g_{WY}, \label{comm} \\
\{ A^X_i,A^Y_j\}_D&=&-\{A_i^X,\phi^{kl}\}(D_1^{-1})_{(kl)(mn)}
\{\psi^{mn},A^Z_r\}\{\tPb^r_Z,A_j^Y\}_D \nonumber \\
&&-\{A_i^X,\tPb^r_Z\}_D\{A_r^Z,\psi^{mn}\}(D_1^{-1})_{(mn)(kl)}
\{\phi^{kl},A^Y_j\}.  \nonumber
\eeq
Here $\Qt_i^X$ is the inverse triad multiplet and 
\be
 I_{(P)X}^Y := \tP^{ i}_X\Pt_i^{Y}, \qquad
 I_{(Q)X}^Y := \tQ^{ i}_X\Qt_i^{Y}
\label{Qi-Qi}
\ee
are projectors on $\tQ$ and $\tP$-multiplets (see Appendix B
for details).   

Quantization may go along the usual way. We may replace the canonical
variables by operators and define a commutator on them
as $[\ . \ ,\ . \ ]:=i\hbar \{ \ . \ ,\ . \ \}_D$. Of course,
when we replace the canonical variables by operators,
the right hand side of (\ref{comm}) becomes ambiguous.
In actual calculations of the area spectrum we will use a shifted
connection $\SA$. As we will see in  section \ref{shiftsec},
for this connection no ordering ambiguity appears.

\section{Area operator and the Wilson line}
\subsection{Wilson line with canonical connection}

In \cite{SA} it was suggested to use the Lorentz covariant
 formulation described above as a basis for 
a modified loop approach. The key point is that $A_i^X$
is a true Lorentz connection (\ref{trans}) and so one can construct
the Wilson line operator 
\be
\widehat U_{\alpha}(a,b)={\cal P}\exp\left(\int_a^b dx^i A_i^X T_X\right), 
\label{hol}
\ee
where $\alpha$ is a path between two points $a$ and $b$,
$T_X$ is a gauge generator.
However, we encounter a serious obstacle 
since instead of 
simple standard
canonical commutation relations now we have a complicated 
algebra of the Dirac brackets (\ref{comm}). 
In particular, the operators like (\ref{hol}) fail to form the loop
algebra. Moreover, since the
connection $A_i^X$ is non-commutative the connection
representation does not exist.
  
Nevertheless, one might hope to obtain some results relying on 
the bracket algebra (\ref{comm}) only. Let us try to 
obtain the spectrum of the area operator extensively investigated
in the framework of the standard loop approach \cite{area,areaAL}.
Here we follow the line of reasonings suggested in \cite{Rov-dif}.
In particular, we use the same regularization technique 
for the area operator. Namely, define the operator of the triad smeared 
over a two-dimensional surface embedded in the 3-manifold:
\be
\tP_X(\Sigma)=\int_{\Sigma} d^2\sigma  \,
n_i(\sigma) \tP_X^i(\sigma),
\ee
where the embedding is described by the coordinates
$x^i(\vec \sigma)$ and
the normal to the surface is given by $n_i=\eps_{ijk}
\frac{\partial x^j}{\partial \sigma^1} 
\frac{\partial x^k}{\partial \sigma^2 }$.
Then the regularized area operator is defined as follows:
\be
{\cal S}=\lim\limits_{\rho \to \infty}\sum\limits_{n} 
\sqrt{g(S_n)}, \label{areaop}
\ee 
where the sum is taken over a partition $\rho$ of $S$ into small
surfaces $S_n$, $\bigcup_n S_n=S$, and
\footnote{Being expressed through $\tPb$ the operator $g(\Sigma)$
reads: $\beta^2 g^{XY}\tPb_X(\Sigma)\tPb_Y(\Sigma)/(\beta^2 -1)$
The printed version of \cite{SA} contains a mistake in
this formula.}
\be
 g(\Sigma)=g^{XY}\tP_X(\Sigma)\tP_Y(\Sigma). \label{3-met}
\ee

We define a state vector corresponding to the Wilson line operator
$\widehat U_{\alpha}$ as
\be
U_\alpha =\widehat U_{\alpha} |0\rangle \,, \label{Ustate}
\ee
where $|0\rangle$ is a vacuum state. 
To be as close as possible to the
connection representation formalism, we require 
\begin{equation}
\tP_X^i |0\rangle =0 \,.
\label{Pvac}
\end{equation}
Since $\tP_X^i$ are commutative, the condition (\ref{Pvac}) is consistent.
The condition (\ref{Pvac}) may lead to troubles if one acts
by the inverse triad on the vacuum state. To avoid problems one
may consider a more general vacuum state with a non-trivial
internal geometry
\begin{equation}
\tP_X^i |0\rangle =\langle\tP_X^i \rangle |0\rangle \,.
\label{Evac}
\end{equation}
Consistency with the second class constraints requires that
$\langle\tP_X^i \rangle$ is expressed through $\langle \tE \rangle$
and $\langle \chi \rangle$ as in (\ref{multHP}).  
After the calculations one can take $\langle\tP_X^i \rangle \to 0$.
The vacuum state (\ref{Evac}) may be also interesting on its own
right (see discussion in section V). 
We shall use primary the simplest vacuum (\ref{Pvac}), but
shall also comment at some points which modifications would
appear if the vacuum (\ref{Evac}) is used instead.

We have constructed a natural generalization of the 
the Wilson line states for the case of non-commutative
Lorentz connection. Let us recall that the
unitary representations of the Lorentz
group are infinite dimensional. Therefore, it is much harder
to address orthogonality, completeness and other functional
properties of the loop states than in the standard $su(2)$
case. We will not discuss these properties here. Instead,
we concentrate on the algebraic aspect of the problem.

To find the area spectrum we study the action of the smeared triad
on a state created by the Wilson line.
Consider the simplest situation when 
the path $\alpha$ has with the surface $\Sigma$
one intersecting point $c$ which breaks $\alpha$ in two parts,
$\alpha_1$ and $\alpha_2$.
Then the action is given by
\begin{eqnarray}
&&\tP_X(\Sigma)\hat U_{\alpha}(a,b)|0\rangle 
=-\int_{\Sigma}d^2\sigma \int_{\alpha}ds\,
\eps_{ijl}\frac{\partial x^i}{\partial \sigma^1} 
\frac{\partial x^j}{\partial \sigma^2 }
\frac{\partial x^k}{\partial s} 
\delta^3({\vec x} (\sigma), {\vec x}(s)) \nonumber \\
&&\hspace{40mm} \times
\hat U_{\alpha_1}(a,c) [ A_k^Y T_Y, \tP^l_X] \hat U_{\alpha_2}(c,b)
|0\rangle \,.\label{PU}
\end{eqnarray}
Here the vacuum state (\ref{Pvac}) has been used. For the vacuum
(\ref{Evac}) an additional term $\langle \tPb_X(\Sigma)\rangle
U_{\alpha}(a,b)$ appears on the right hand side of (\ref{PU}).

In the standard loop approach \cite{area,areaAL} one has to
consider the action of the smeared triad $\tE$ on the Wilson
line with $su(2)$ connection $A_i^a$. Therefore,
the equation (\ref{PU}) should be replaces by an analogous one with
the commutator of the canonical variables $[A_i^a,\tE_b^j]$
on the right hand side. This commutator is proportional to
$\delta_i^j$. Because of this fact, 
the explicit $x$-dependence can be canceled, and 
the right hand side of 
$\tE U_\alpha$ becomes in the standard loop approach a
purely algebraic expression. As a result the area operator (that is
essentially $\tE$ applied twice) can be easily diagonalized.
In the present case
$\{ A_k^Y , \tP^l_X\}_D$ is {\it not} proportional to $\delta^l_k$.
Consequently, the area operator acting on the Wilson line
$U_\alpha$ with the canonical connection $A$
is not just a matrix in the Lorentz indices and cannot be that
easily made diagonal. A way to by-pass this difficulty is suggested
in the next section \ref{shiftsec}.

\subsection{Shifted connection}\label{shiftsec}
We have seen that to enable  diagonalization the area operator
the commutator of the connection and $P$ should be unit matrix
in the spatial indices. It is known that if one adds a vector
to a connection the resulting object will again transform
as a connection. We are going to use this arbitrariness in the
choice of the connection to diagonalize the area operator.
We are interested in a new connection $\SA_i^X$ such that:
i) it is a true Lorentz connection, i.e. $\SA_i^X -A_i^X$
is tensorial in both indices;
 ii) the Dirac bracket
$\{ \SA_k^Y , \tP^l_X\}_D$ is proportional to $\delta^l_k$;
iii) $\SA_i^X-A_i^X$ is proportional to the first class constraints.
These requirements appear to be very strong.
There is just one connection which satisfies all of them.
To show this, let us note that all the triad (or tetrad)
components have dimension zero, while the connection has
mass dimension one. Consequently the Gauss constraint has
dimension one, and the diffeomorphism and Hamiltonian 
constraints have dimension two. It is clear therefore
that
\be
\SA_i^X=A_i^X+\alpha^{XY}_i(Q) \G_Y\,,
\label{newcon0} 
\ee
where $\alpha^{XY}_i(Q)$ does not contain derivatives or connections.
The coefficient fucntions $\alpha^{XY}_i(Q)$ have to be
tensorial in order to ensure 
correct diffeomorphism and Lorentz transformation properties
of $\SA$:
\beq
\{ \G_X,\SA^Y_i\}_D&=&
\delta^Y_X\partial_i -f_{XZ}^Y \SA^Z_i,  \label{newtrans} \\
\{ {\cal D}(\vec N), \SA_i^X \}_D&=&
\SA_j^X\partial_i N^j+N^j\partial_j \SA_i^X. \label{newAdiff}
\eeq
${\cal D}(\vec N)$ is defined in Appendix C (\ref{smcon}).
Thus $\SA_i^X$ is the true so(3,1) connection. 
There is still a 6-parameter family of the
connections which satisfy (\ref{newtrans}) and (\ref{newAdiff}).
This ambiguity is fixed  uniquely by the 
second condition ii). We arrive at  the following
Lorentz connection: 
\be
\SA_i^X=A_i^X+\frac{1}{2\left(1+\frac{1}{\beta^2}\right)}
\R^{X}_{S}I_{(Q)}^{ST}\R_T^Z f^Y_{ZW}\Pt_i^W \G_Y.
\label{newcon} 
\ee

The connection $\SA_i^X$ has a very simple bracket with $\tP_Y^j$
\be 
\{ \SA^X_i,\tP_Y^j\}_D=\delta_i^j I_{(P)Y}^X. \label{newAP} 
\ee 
Already at this point we observe independence of the right hand
side of (\ref{newAP}) from $\beta$. It should be stressed that
this $\beta$-independence is {\it not} a pre-requirement
in our construction. This is rather a consequence of the
conditions i)--iii) above. We observe also
\beq
\{ \SA^X_i,\Pt_j^Y\}_D&=&-\Pt_j^X\Pt_i^Y, \label{newAiP} \\
\{ \SA^X_i,I_{(P)}^{YZ}\}_D&=&0. \label{newAI} 
\eeq
Due to this relation the projectors $I_{(P)}$ and $I_{(Q)}$ behave 
very similar to $c$-numbers.

The Dirac bracket of two connections has a very complicated form and
will not be presented here. However, an important observation
can be made already by considering 
the Jacobi identity 
\be 
\{ \{ \SA_i^X, \SA_j^Y\}_D , \tP^k_Z \}_D=
\{ \{ \SA_i^X, \tP^k_Z\}_D , \SA^Y_j \}_D -
\{ \{ \SA_j^Y, \tP^k_Z\}_D , \SA_i^X \}_D=0. \label{Jac}
\ee
It follows from (\ref{Jac}) that $\{ \SA_i^X, \SA_j^Y\}_D$ 
does not depend on the
connection. It is a function of $\tQ$ and its derivatives, i.e.
this bracket contains only commuting objects on the right hand side.
Therefore, there will be no ordering ambiguity if we replace
the Dirac brackets with $\SA_i^X$ by the corresponding operator relation.
We will use this as a new quantization rule. In particular,
\be
[\SA_i^X,\tP_Y^j]=i\hbar \delta_i^j I_{(P)Y}^X \,.
\label{newquant}
\ee

Note, that the commutators with the new connection (\ref{newcon})
are insufficient to define all commutators involving the canonical
connection. The reason is that the (classical) field $\SA_i^X$
satisfies the condition
\be
g_{YZ}(\delta_i^k I_{(Q)X}^{Y}-\Qt_i^Y \tQ^k_X)\SA_k^Z=
I_{(Q)X}^Y f^W_{YZ} \Qt_i^Z \partial_j \tQ^j_W 
\ee
and has fewer independent components than $A_i^X$. From  
(\ref{newcon}) it is clear that the
missing components are contained in the Gauss constraint.
For practical purposes it is therefore enough to know
the commutators with $\SA_i^X$ {\it and} the commutators with
the Gauss constraint which are defined either by the structure constants
of the Lorentz group or by the matrix elements in corresponding
representations. These quantization rules have one more
important advantage. They ensure that quantum transformation laws 
are identical to the classical ones. So
there will be no gauge anomaly for the
Lorentz group.

\section{Area spectrum}
The shifted connection $\SA$ can be used as an argument 
the Wilson line. 
Let us evaluate action of the area operator (\ref{areaop})
on the states created by such Wilson lines. It is given by
\begin{equation}
{\cal S}\hat U_\alpha [\SA ]|0\rangle =\hbar \hat U_{\alpha_1}[\SA ] 
\sqrt{-I_{(P)}^{XY}T_XT_Y} \hat U_{\alpha_2}[\SA ]
|0\rangle \,,\label{au}
\end{equation}
where we used equations (\ref{PU}) and (\ref{newquant}) and the
prescription \cite{area,areaAL} for taking the square root
of the operator (assuming that the latter is still valid for
the Lorentz gauge group). Vacuum state is supposed to be the
trivial one (\ref{Pvac}).

Consider the matrix operator $I_{(P)}^{XY}T_XT_Y$. It can be rewritten
as
\begin{equation}
I_{(P)}^{XY}T_XT_Y =g^{XY}T_XT_Y -I_{(Q)}^{XY}T_XT_Y \,,
\label{PgQ}
\end{equation}
where the first term is a quadratic Casimir of the Lorentz
algebra:
\begin{equation}
g^{XY}T_XT_Y =C_2(so(3,1)) \,.\label{LCas}
\end{equation}
In order to study the second term in (\ref{PgQ}) let us introduce
the generators
\begin{equation}
q_a:=\frac{1}{\sqrt{1-\chi^2}}\left(\delta_{ab}-
\frac{1-\sqrt{1-\chi^2}}{\chi^2}\chi_a\chi_b\right)
\Et_i^b \tQ^i_X T^X\, .
\label{qa}
\end{equation}
One can check directly that 
\beq
&  I_{(Q)}^{XY}T_XT_Y=-q_aq_a, & \\ 
& [q_a,q_b]=-{\eps_{ab}}^cq_c \,. & \label{qcom}
\eeq
Consequently, $q_a$ generate the $so(3)$ subalgebra  of $so(3,1)$, and 
$I_{(Q)}^{XY}T_XT_Y$ is the Casimir operator of this subalgebra
\begin{equation}
q_aq_a=-C_2(so(3)).
\label{Crot}
\end{equation}
In a suitable basis in the defining representation of $so(3,1)$
the generators $q_a$ annihilate the vector 
$v_\chi =(1-\chi^2)^{-1/2}(1,\chi_a)$. All vectors $v_\chi$ belong
to the same orbit of the Lorentz group. Therefore, the subalgebras
spanned by $\{ q_a \}$ for different $\chi$ are conjugate in $so(3,1)$,
and spectrum of $so(3)$ representations obtained 
after the restriction $so(3,1)\downarrow so(3)$
from a given representation of $so(3,1)$ does not depend on $\chi$.
Eigenvalues of the Casimir operator (\ref{Crot})
are also $\chi$-independent.

Spectrum of the area operator acting on Wilson lines reads:
\begin{equation}
{\cal S}\sim \hbar \sqrt{-C_2(so(3,1))+C_2(so(3))}
\,.\label{areaspec}
\end{equation}
This formula represents the main result of our paper. 

One can think naively that the Lorentz invariance of
the area spectrum (\ref{areaspec}) is broken
due to the presence of the Casimir operator of a subgroup.
This is however not the case.
Under local Lorentz transformations the Wilson line
changes as $U(x,y)\to {\cal U}(x) U(x,y) {\cal U}^{-1}(y)$,
where ${\cal U}(x)$ is an element of the Lorentz group
taken in an appropriate representation. The matrix operator
$\sqrt{-I_{(P)}^{XY}T_XT_Y}$ changes in a similar way:
$\sqrt{-I_{(P)}^{XY}T_XT_Y}\to {\cal U}(x)
\sqrt{-I_{(P)}^{XY}T_XT_Y}{\cal U}^{-1}(x)$. Thus proper
(covariant) transformation properties of (\ref{au}) are
recovered.

As expected, the area spectrum (\ref{areaspec}) does
not depend on the Immirzi parameter $\beta$.

\section{Discussion}
In this paper we analysed the area operator spectrum in
a manifestly Lorentz covariant formalism. We have constructed
a generalization of the Wilson line states for the
case of non-commutative connection. As usual, there is
certain arbitrariness in the choice of the connection.
Namely, any connection can be shifted by a vector and
would still remain a connection. We use this arbitrariness to
define a connection $\SA$ such that 
$\{ \SA_i^X,\tP_Y^j\}_D\sim \delta_i^j$. 
Because of 
the rather simple commutation relations (\ref{newquant})
we are able to find explicitly the area spectrum 
(\ref{areaspec}). Since the right hand side of (\ref{newquant})
does not depend on the Immirzi parameter $\beta$, there is
no dependence on $\beta$ in the spectrum (\ref{areaspec}) as
well.

Note, that the connection $\SA$ is unique only if we require that
it coincides with $A$ on the surface of the constraints.
A different idea might be to fix the connection by considering
its space-time properties. Because of the rather complicated
form of the Dirac brackets with the Hamiltonian constraint
this is a technically very involved calculation. We may hope
that the results obtained in this way will agree with
our results.

We must admit that there is no proof in this paper that
the area spectrum with {\it any} connection does not
depend on $\beta$. We cannot perform direct
calculations with a connection other than $\SA$.
We may, however, {\it interpret} the shift $A \to \SA$
as diagonalization of the area operator. Our results
suggest that in a Lorentz covariant quantization the
dependence of the physical quantities on the Immirzi
parameter ultimately disappears.

In addition to the explicit Lorentz covariance there is
another advantage of our approach. The Hamiltonian constraint
(\ref{Sd3}) is polynomial in the canonical variables
(as for the Ashtekar or Euclidean cases). 
Due to this the corresponding
regularized quantum operator may be similar to the first term of 
Thiemann's constraint operator \cite{Tim}.
That would eliminate difficulties created by the second term.
Note that the spin foam formulation of loop quantum gravity
takes into account the first term of Thiemann's
Hamiltonian only \cite{Rov-dif,sf}.

Let us comment on the choice of the vacuum state.
The connection representation implies that the trivial
vacuum (\ref{Pvac}) is chosen. Such representation does
not exist in our case due to the non-commutativity of the
connection fields. Therefore, we must choose a vacuum state
explicitly.
The possibility of a more general vacuum state 
(\ref{Evac}) can be taken into account.
(A similar possibility has been already discussed in \cite{BLM}).
For the vacuum state (\ref{Evac}) we have no problem
with the action of the inverse triad on the vacuum, but
we loose explicit background independence. Physical consequences
of different vacua have to be clarified yet.

Even without relation to the Immirzi parameter problem
quantization of gravity in manifestly Lorentz invariant
terms is an important task. We have considered here the
algebraic part of the problem, while the functional analysis
part has been completely ignored. We do not know how to
construct a complete orthogonal basis in the space of
states out of the Wilson lines. Consequently, we may only
guess which representations do actually contribute to
the area spectrum (\ref{areaspec}).

The area spectrum (\ref{areaspec}) now contains the Casimir
operator of non-compact Lorentz group. Since unitary representations
of the Lorentz group are labelled by a pare of indices $(\rho ,j)$,
and the index $\rho$ is continuous, we may expect that the area
spectrum becomes continuous as well. This would be a new feature
for the loop quantum gravity, though continuous spectrum appears
in the spin foam models \cite{sfLor}. However, in the view of the
remarks in the previous paragraph, this feature should be taken
with great amount of care. 

Recently, a manifestly  
$so(3,1)$-covariant formalism has been developed in the 
framework of spin foam models 
\cite{sfLor}. It has been suggested to use the so-called 
simple representations of the Lorentz group only. The Immirzi
parameter has been also included in this approach
\cite{Mon,Liv}. The area spectrum obtained in the spin foam
models is different from our expression (\ref{areaspec}).
The reason is that we use different quantization rules.
We should stress that our commutation relations are
{\it derived} from the gravitational action rather than
postulated.
Therefore, our quantization rules may provide a more solid
ground for the Lorentz-invariant spin foam models.
Despite of complicated Dirac brackets our final
commutation relations (\ref{newquant}) are rather simple.
It should be possible to use them in the spin foam approach.

\section*{Acknowledgements}
We are grateful to Abhay Ashtekar for fruitful discussions.
This work of D.V. has been partially supported by the DFG project
Bo 1112/11-1. The research of S.A. has been supported in part by 
European network EUROGRID HPRN-CT-1999-00161.

\appendix

\section{Matrix algebra}

In the basis (\ref{multHP}) 
the $so(3,1)$ structure constants are:
\be
\begin{array}{ccc}
f_{A_1 A_2}^{A_3}=0,&
f_{A_1 B_2}^{A_3}=-\eps^{A_1 B_2 A_3},&
f_{B_1 B_2}^{A_3}=0, \\
f_{B_1 B_2}^{B_3}=-\eps^{B_1 B_2 B_3},&
f_{A_1 B_2}^{B_3}=0,&
f_{A_1 A_2}^{B_3}=\eps^{A_1 A_2 B_3}.
\end{array}      \label{algHP}
\ee
Here we split the 6-dimensional index $X$ into a pair of 3-dimensional
indices, $X=(A,B)$, so that $A,B=1,2,3$. $\eps$ is the Levi--Civita
symbol, $\eps^{123}=1$.

All triad multiplets are connected by numerical matrices:
\beq
\tP^i_X=\Pi_X^Y\tQ^i_Y,&\qquad &
\Pi^{Y}_X =\left(
\begin{array}{cc}
0&1 \\ -1&0
\end{array}
\right)\delta_a^b,   \label{p-q}     \\
\tP^i_X=\frac{\R_X^Y}{1+\frac{1}{\beta^2}}\tPb^i_Y,&\qquad &
\R^{Y}_X =\left(
\begin{array}{cc}
1& -\frac{1}{\beta} \\ 
 \frac{1}{\beta}&1
\end{array}
\right)\delta_a^b.     \label{pb-q}
\eeq
They as well as their inverse commute with each other and, furthermore, 
they commute with the structure constants
in the following sense:
\be f^{XYZ'}\Pi_{Z'}^Z=f^{XY'Z}\Pi_{Y'}^Y. \label{com-f} \ee
Other useful relations can be found in \cite{SA}.

\section{Inverse multiplets and projectors}

The inverse triad multiplets are introduced as the following fields:
\beq
\Pt_i^{X}&=&\left( \frac{\delta^a_b-\chi^a\chi_b}{1-\chi^2}\Et_i^b,
-\frac{ {\eps^a}_{bc}\Et_i^b\chi^c}{1-\chi^2} \right),
\nonumber \\
\Qt_i^{X}&=&\left( \frac{ {\eps^a}_{bc}\Et_i^b\chi^c}{1-\chi^2},
\frac{\delta^a_b-\chi^a\chi_b}{1-\chi^2}\Et_i^b \right).
\label{invQ}\eeq
They satisfy:
\be
 \{ \G_X,\Pt_i^Y\}=-f_{XZ}^Y\Pt_i^Z, \quad
 \tP^i_X\Pt_j^X=\delta^i_j, \quad
\tQ^i_X\Pt_j^X=0.  \nonumber
\ee
Similar properties are valid for $\Qt_i^X=\Pi^X_Y\Pt_i^Y$.

The projectors (\ref{Qi-Qi}) read:
\begin{equation}
 I_{(P)X}^Y =\left(
\begin{array}{cc}
\frac{\delta_a^b-\chi_a\chi^b}{1-\chi^2} &
\frac{ {\eps_a}^{bc}\chi_c}{1-\chi^2} \\
\frac{ {\eps_a}^{bc}\chi_c}{1-\chi^2} &
-\frac{\delta_a^b\chi^2-\chi_a\chi^b}{1-\chi^2}
\end{array} \right) \label{Qi-Qi-exact}
\end{equation}
and  $I_{(Q)X}^Y=\delta^Y_X -I_{(P)X}^Y$.
Besides, one can note the relations which are very helpful in
calculations:
\beq 
& I_{(P)}^{XY}=-\Pi^X_Z I_{(Q)}^{ZW} \Pi_W^Y, & \\
&  f^{WYZ} I_{(P)W}^X \tQ^i_Y \tQ^j_Z =0, & \\
& f^{WYZ} I_{(Q)W}^X \tQ^i_Y \tQ^j_Z = f^{XYZ} \tQ^i_Y \tQ^j_Z. & 
\eeq

The commutators of the second class constraints
form the following triangular matrix:
\be \Delta=\left(
\begin{array}{cc}
0 & D_1 \\
-D_1 & D_2
\end{array} \right), \quad
\Delta^{-1}=\left(
\begin{array}{cc}
D_1^{-1} D_2 D_1^{-1} & -D_1^{-1} \\
D_1^{-1} & 0
\end{array}  \right),   \label{Delta}
\ee
where
\beq
D_1^{(ij)(kl)}&=&
\{ \phi^{ij}, \psi^{kl}\}=\frac{4\beta^2}{1+\beta^2}(\tQ\tQ)^{\{i[j\}}
(\tQ\tQ)^{\{k]l\}} , \label{D1} \\
(D_1^{-1})_{(kl)(mn)}&=&
\frac{1}{8}\left(1+\frac{1}{\beta^2}\right)
\left( (\Qt\Qt)_{kl}(\Qt\Qt)_{mn}-(\Qt\Qt)_{km}(\Qt\Qt)_{ln}-
(\Qt\Qt)_{kn}(\Qt\Qt)_{lm} \right).   \label{D-1}
\eeq
Explicit form of $D_2$ is not 
needed since all brackets are expressed in terms of $D_1^{-1}$
only (see (\ref{comm})).

\section{Constraint algebra}
Define the smeared constraints:
\begin{eqnarray}
&&{\cal G}(n)=\int d^3x\, n^X{\cal G}_X, \qquad
H(\tN )=\int d^3x\, \tN H,
\nonumber \\
&&{\cal D}(\vec N)=\int d^3x\, N^i(H_i+A_i^X{\cal G}_X). \label{smcon}
\end{eqnarray}
They obey the following algebra:
\begin{eqnarray}
&&\left\{ {\cal G}(n) ,{\cal G}(m) \right\}_D={\cal G}(n\times m),
\nonumber \\
&&\left\{ {\cal D}(\vec N) ,{\cal D}(\vec M) \right\}_D=
-{\cal D}([\vec N ,\vec M ]),\nonumber \\
&&\left\{ {\cal D}(\vec N) ,{\cal G}(n) \right\}_D=-
{\cal G}( N^i\partial_in), \nonumber \\
&&\Bigl\{ H(\tN ) ,{\cal G}(n) \Bigr\}_D =0, \label{algA} \\
&&\Bigl\{ {\cal D}(\vec N) ,H(\tN ) \Bigr\}_D=
-H({\cal L}_{\vec N}\tN ), \nonumber \\
&&\Bigl\{ H(\tN ),H(\tM ) \Bigr\}_D =
{\cal D}(\vec K)-{\cal G}(K^jA_j), \nonumber
 \end{eqnarray}
where
\begin{eqnarray}
&&(n\times m)^X=f^X_{YZ}n^Ym^Z,\qquad
{\cal L}_{\vec N}\tN =
N^i\partial_i \tN-\tN\partial_iN^i, \nonumber \\
&&[\vec N ,\vec M ]^i=
N^k\partial_kM^i-M^k\partial_kN_i, \label{not1} \\
&& K^j=(\tN\partial_i\tM-\tM\partial_i\tN)\tQ^i_X\tQ^j_Y g^{XY}.
\nonumber
\end{eqnarray}

\end{document}